\documentclass[12pt,a4paper]{article} 
\usepackage[dvips]{graphicx} 
\begin{document}
\begin{titlepage}
\begin{flushright}
UAB--FT--525\\
hep-ph/0204339\\
April 2002
\end{flushright}
\vspace*{1.6cm}

\begin{center}
{\Large\bf 
Scalar $f_0(980)$ and $\sigma (500)$ meson exchange\\ in $\phi$ decays 
into $\pi^0\pi^0\gamma$}\\
\vspace*{0.8cm}

A.~Bramon$^1$, R.~Escribano$^{1,2}$, J.L.~Lucio M.$^3$, M.~Napsuciale$^3$ 
and G.~Pancheri$^4$\\
\vspace*{0.2cm}

{\footnotesize\it
$^1$Grup de F\'{\i}sica Te\`orica,
Universitat Aut\`onoma de Barcelona, E-08193 Bellaterra (Barcelona), Spain\\
$^2$IFAE, 
Universitat Aut\`onoma de Barcelona, E-08193 Bellaterra (Barcelona), Spain\\
$^3$Instituto de F\'{\i}sica, Universidad de Guanajuato,
Lomas del Bosque \#103, Lomas del Campestre, 37150 Le\'on, Guanajuato, Mexico\\
$^4$INFN-Laboratori Nazionali di Frascati, P.O.~Box 13, I-00044 Frascati, Italy.}

\end{center}
\vspace*{0.2cm}

\begin{abstract}
The complementarity between Chiral Perturbation Theory and the Linear Sigma Model 
is exploited to study $\pi^0\pi^0$ production in $\phi$ radiative decays,
where the effects of the $f_0(980)$ scalar resonance, and those of its more 
controversial $\sigma(500)$ partner, should manifest via the 
$\phi\rightarrow K^+ K^- (\gamma)\rightarrow\pi^0\pi^0\gamma $ decay chain. 
The recently reported data on $\phi\rightarrow\pi^0\pi^0\gamma$ 
coming from the VEPP-2M $e^+ e^-$ collider in Novosibirsk and the
DA$\Phi$NE $\phi$-factory in Frascati
can be reasonably described in our approach, which we propose as a promising 
first step towards more detailed analyses.  
The $f_0(980)$ contribution,
which appears as a moderately narrow peak at the high part of the dipion mass spectrum,
can be interpreted as the isoscalar member of the scalar nonet with a large
$f_0 K\bar{K}$ coupling and an $f_0\pi\pi$ coupling suppressed by almost ideal 
$\sigma$-$f_0$ mixing. 
Indeed, the mixing angle in the flavour basis is found to be $\phi_{S}\approx -6^\circ$,
if the $f_0$-propagator is approximated by a simple Breit-Wigner,
or $\phi_{S}\approx -9^\circ$, if an improved two-channel analysis is performed.
The $\sigma(500)$ resonance, which is then strongly coupled to pion pairs,
yields a tiny contribution  because, in our approach, its coupling to kaon pairs 
is proportional to $m^2_\sigma-m^2_K$ and thus quite small.     
\end{abstract}
\end{titlepage}

\section{Introduction}
The radiative decays of low mass vector mesons into two neutral pseudoscalars,
$V\rightarrow P^0 P^0 \gamma$, are known to be a useful tool to investigate the
complicated dynamics governing meson physics around 1 GeV.   
Particularly interesting are those decays proceeding mainly by the exchange of scalar 
resonances because of the unconventional \cite{Jaffe:1976ig} and controversial nature
of these states, as exemplified, for instance, in the very recent analysis by
Close and T\"{o}rnqvist \cite{Close:2002zu}.
Experimental data on the $\phi\rightarrow\pi^0\pi^0\gamma$ decays where, after the
emission of a photon, a $K^+K^-$ pair rescatters into $\pi^0\pi^0$ in a process dominated
by the exchange of isoscalar scalar resonances in the $s$-channel
have thus been watched for with great interest \cite{Close:ay,Achasov:1987ts}. 

The first measurements of this $\phi\rightarrow\pi^0\pi^0\gamma$ decay have been reported
by the SND and CMD-2 Collaborations.  
For the branching ratio, they obtain   
\begin{equation}
\label{VEPP} 
B(\phi\rightarrow\pi^0\pi^0\gamma)=\left\{
\begin{array}{ll}
(1.22\pm 0.10\pm 0.06)\times 10^{-4} & \mbox{SND   \protect\cite{Achasov:2000ym}}\\[1ex]
(0.92\pm 0.08\pm 0.06)\times 10^{-4} & \mbox{CMD-2 \protect\cite{Akhmetshin:1999di}}
\end{array}
\right.
\end{equation}
for $m_{\pi^0\pi^0}>700$ MeV in the latter case. 
More recently, the KLOE Collaboration has measured \cite{Aloisio:2002bt}
\begin{equation}
\label{KLOE}
B(\phi\rightarrow\pi^0\pi^0\gamma)=(1.09\pm 0.03\pm 0.05)\times 10^{-4}\ ,
\end{equation}
in agreement with (\ref{VEPP}) but with a considerably smaller error.
In all the cases, the spectrum is clearly peaked at $m_{\pi^0\pi^0}\simeq 970$ MeV, 
as expected from an important $f_0(980)$ contribution.
This and other radiative vector meson decays are now being further investigated 
at the Frascati $\phi$-factory DA$\Phi$NE \cite{daphne95:franzini}
with higher accuracy (see for example Ref.~\cite{Aloisio:2002bs}).

On the theoretical side, the $\phi\rightarrow \pi^0\pi^0\gamma$ decay has been 
considered by a number of authors \cite{Bramon:1992kr}--\cite{Gokalp:2001bk}.
Early calculations of the vector meson dominance (VMD) amplitude for these processes,
{\it i.e.}~the contributions proceeding through the decay chain 
$\phi\rightarrow\pi^0\rho^0\rightarrow\pi^0\pi^0\gamma$,
were summarized in Ref.~\cite{Bramon:1992kr}.
The $\phi$-decay vertex is forbidden by the Zweig rule and the width and branching ratios
predicted by VMD, $\Gamma(\phi\rightarrow\pi^0\pi^0\gamma)_{\rm VMD}= 51$ eV and   
$B(\phi\rightarrow\pi^0\pi^0\gamma)_{\rm VMD}=0.12\times 10^{-4}$ \cite{Bramon:1992kr},    
are found to be substantially smaller than the experimental results quoted 
in Eqs.~(\ref{VEPP}) and (\ref{KLOE}).
Moreover the VMD spectrum peaks at low $m_{\pi\pi}$ values well below the $f_0(980)$ 
resonance mass.

The possibility of an enhancement in this branching ratio through the
$\phi\rightarrow K^+K^-(\gamma )\rightarrow\pi^0\pi^0\gamma$ mechanism 
\cite{Close:ay,Achasov:1987ts} was pointed out in
Ref.~\cite{Bramon:1992kr} and further discussed in Ref.~\cite{Bramon:1992ki}  
in a Chiral Perturbation Theory (ChPT) context enlarged to include on-shell vector mesons. 
This formalism gives well defined predictions for the various
$V\rightarrow P^0P^0\gamma$ decays in terms of $P^+P^-\rightarrow P^0P^0$
rescattering amplitudes, which are easily calculated in strict ChPT,
and a loop integral over the intermediate $P^+P^-$ pair.
In this approach, the $\phi\rightarrow\pi^0\pi^0\gamma$ decay is dominated by
(Zweig rule allowed) kaon loops leading to
$\Gamma(\phi\rightarrow\pi^0\pi^0\gamma)_{\chi}= 224$ eV,  
while (Zweig rule forbidden) pion loop contributions are strongly suppressed.  
The corresponding spectrum peaks at moderate values of $m_{\pi\pi}$ but 
the interference between this kaon loop contribution and the previous 
VMD amplitude turns out to be quite small leading globally to  
$\Gamma(\phi\rightarrow\pi^0\pi^0\gamma)_{{\rm VMD}+\chi}=269$ eV and
$B(\phi\rightarrow\pi^0\pi^0\gamma)_{{\rm VMD}+\chi}=0.61\times 10^{-4}$ \cite{Bramon:1992ki}, 
which are still below the experimental results quoted in Eqs.~(\ref{VEPP}) 
and (\ref{KLOE}). 
Additional contributions are thus certainly required. 
The most natural candidates for closing this gap between theory and experiment 
are the contributions coming from the exchange of scalar resonances, such as 
the well established $f_0(980)$ and the more controversial $\sigma(500)$ 
(or $f_0$(400--1200)) mesons \cite{Groom:2000in}.
 
The purpose of this paper is to study the effects of the low mass scalar states in the 
$\phi\rightarrow\pi^0\pi^0\gamma$ decays following the ChPT inspired context
introduced in Refs.~\cite{Bramon:2000vu,Escribano:2000fs,Bramon:2001un}
to account similarly for the scalar exchange contributions to
$\phi\rightarrow\pi^0\eta\gamma$ and $\rho,\omega\rightarrow\pi^0\pi^0\gamma$.  
In this context, one takes advantage of the common origin of ChPT and the L$\sigma$M
to improve the chiral loop predictions for $V\rightarrow P^0P^0\gamma$ exploiting 
the complementarity of both approaches for these specific processes.
As a result, a simple analytic amplitude, 
${\cal A}(\phi\rightarrow\pi^0\pi^0\gamma )_{\mbox{\scriptsize L$\sigma$M}}$, 
will be obtained which includes the effects of the scalar meson poles and also shows 
the appropriate behaviour expected from ChPT at low dipion invariant masses. 
There also exists the contribution to $\phi\rightarrow\pi^0\pi^0\gamma$ 
coming from the previously mentioned vector meson exchange.
This VMD amplitude, 
${\cal A}(\phi\rightarrow\pi^0\pi^0\gamma )_{\mbox{\scriptsize VMD}}$, 
is well known and scarcely interesting but has to be added to  
${\cal A}(\phi\rightarrow\pi^0\pi^0\gamma )_{\mbox{\scriptsize L$\sigma$M}}$, 
{\it i.e.}~to the relevant amplitudes containing the scalar meson effects, 
in order to compare with available and forthcoming data.
We will conclude that data on the $\phi\rightarrow\pi^0\pi^0\gamma$ channel can 
contribute decisively to improve our knowledge on the $f_0(980)$ scalar resonance, 
as well as on some specific features of its elusive and controversial partner
$\sigma(500)$. 
 
\section{Chiral loop contributions to $\phi\rightarrow\pi^0\pi^0\gamma$}
\label{sectChPT}
The vector meson initiated $V\rightarrow P^0 P^0\gamma$ decays cannot be treated in 
strict Chiral Perturbation Theory (ChPT). 
This theory has to be extended to incorporate on-shell vector meson fields.
At lowest order, this may be easily achieved by means of the ${\cal O}(p^2)$ 
ChPT Lagrangian
\begin{equation}
\label{ChPTlag}
{\cal L}_2=\frac{f^2}{4}\langle D_\mu U^\dagger D^\mu U+M(U+U^\dagger )\rangle\ ,
\end{equation}
where $U=\exp(i\sqrt{2}P/f)$, $P$ is the usual pseudoscalar octet matrix and, 
at this order,
$M=\mbox{diag}(m_\pi^2, m_\pi^2, 2m_K^2-m_\pi^2)$ and $f=f_\pi=92.4$ MeV.
The covariant derivative, now enlarged to include vector mesons, is defined as 
$D_\mu U=\partial_\mu U -i e A_\mu [Q,U] - i g [V_\mu,U]$ with 
$Q=\mbox{diag}(2/3, -1/3, -1/3)$ being the quark charge matrix and $V_\mu$ 
the additional matrix containing the nonet of vector meson fields. 
We follow the conventional normalization for the vector nonet matrix such that the 
diagonal elements are $(\rho^0+\omega_0)/\sqrt{2}, (-\rho^0+\omega_0)/\sqrt{2}$ and
$\phi_0$, where $\omega_0 = (u\bar{u}+d\bar{d})/\sqrt 2$ and $\phi_0 = s\bar{s}$
stand for the ideally mixed states.
The physical $\phi$-field is approximately $\phi_0+\epsilon\omega_0$,
where $\epsilon\simeq +0.059\pm 0.004$ accounts for the $\omega$-$\phi$ mixing angle
in the flavour basis \cite{Groom:2000in}.

One easily observes that there is no tree-level contribution from the Lagrangian 
(\ref{ChPTlag}) to the $\phi\rightarrow\pi^0\pi^0\gamma$ amplitude and that,
at the one-loop level, one needs to compute the set of diagrams shown in 
Ref.~\cite{Bramon:1992ki}. 
We do not take into account pion loop contributions here since they proceed only through
the small $\epsilon$ piece of the $\phi$-field (because of the Zweig rule) which is further
suppressed by $G$-parity, as in $\omega\rightarrow\pi^+\pi^-$.
These pion loop contributions are known to be negligible as compared to those from kaon loops
proceeding through the Zweig rule allowed piece, $\phi_0=s\bar{s}$, not restricted by
$G$-parity  arguments \cite{Bramon:1992ki}. 

A straightforward calculation leads to the following {\it finite} amplitude for 
$\phi(q^\ast,\epsilon^\ast)\rightarrow\pi^0(p)\pi^0(p^\prime)\gamma(q,\epsilon)$
(see Ref.~\cite{Bramon:1992ki} for further details):
\begin{equation}
\label{AphiChPT}
{\cal A}(\phi\rightarrow\pi^0\pi^0\gamma)_\chi= 
\frac{eg_{s}}{2\pi^2 m^2_{ K^+}}\,\{a\}\,L(m^2_{\pi^0\pi^0})\times
{\cal A}(K^+ K^-\rightarrow\pi^0\pi^0)_\chi\ ,
\end{equation} 
where
$\{a\}=(\epsilon^\ast\cdot\epsilon)\,(q^\ast\cdot q)-
       (\epsilon^\ast\cdot q)\,(\epsilon\cdot q^\ast)$, 
$m^2_{\pi^0\pi^0}\equiv s\equiv (p+p^\prime)^2=(q^\ast -q)^2$ is the invariant mass
of the final dipion system and $L(m^2_{\pi^0\pi^0})$ is the loop integral function
defined as \cite{Close:ay,Achasov:1987ts,Bramon:1992ki,Marco:1999df,LucioMartinez:uw}
\begin{equation}
\label{L}
\begin{array}{rl}
L(m^2_{\pi^0\pi^0}) &=\ 
\frac{1}{2(a-b)}-
\frac{2}{(a-b)^2}\left[f\left(\frac{1}{b}\right)-f\left(\frac{1}{a}\right)\right]\\[2ex]
&+\ \frac{a}{(a-b)^2}\left[g\left(\frac{1}{b}\right)-g\left(\frac{1}{a}\right)\right]\ 
\end{array}
\end{equation}
with 
\begin{equation}
\label{f&g}
\begin{array}{l}
f(z)=\left\{
\begin{array}{ll}
-\left[\arcsin\left(\frac{1}{2\sqrt{z}}\right)\right]^2 & z>\frac{1}{4}\\[1ex]
\frac{1}{4}\left(\log\frac{\eta_+}{\eta_-}-i\pi\right)^2 & z<\frac{1}{4}
\end{array}
\right.\\[5ex]
g(z)=\left\{
\begin{array}{ll}
\sqrt{4z-1}\arcsin\left(\frac{1}{2\sqrt{z}}\right) & z>\frac{1}{4}\\[1ex]
\frac{1}{2}\sqrt{1-4z}\left(\log\frac{\eta_+}{\eta_-}-i\pi\right) & z<\frac{1}{4}
\end{array}
\right.
\end{array}
\end{equation}
and
$\eta_\pm=\frac{1}{2}(1\pm\sqrt{1-4z})$, $a=m^2_\phi/m^2_{K^+}$ and 
$b=m^2_{\pi^0\pi^0}/m^2_{K^+}$.
The coupling constant $g_{s}$ comes from the strong decay amplitude
${\cal A}(\phi\rightarrow K^+K^-)=g_{s}\,\epsilon^\ast\cdot (p_+-p_-)$
and takes the value $|g_s|\simeq 4.6$ to agree with
$\Gamma (\phi\rightarrow K^+K^-)_{\rm exp}= 2.19$ MeV \cite{Groom:2000in}. 
This coupling is the part beyond standard ChPT which we have fixed
phenomenologically.

The four-pseudoscalar amplitude is instead a standard ChPT amplitude
which is found to depend linearly on the variable $s=m^2_{\pi^0\pi^0}$ only:
\begin{equation}
\label{A4PChPTphys}
{\cal A}(K^+K^-\rightarrow\pi^0\pi^0)_\chi=\frac{s}{4 f_\pi f_K}\ ,
\end{equation}
with $f_K = 1.22 f_\pi$. 
It is important to notice that in deriving this ChPT amplitude no use has been made of
the relation $s+t+u=2m^2_K+2m^2_\pi$ valid only for on-shell particles but not in our case
with off-shell kaons in the chiral loop. 
Thanks to this genuine $t$ and $u$ independence, the chiral amplitude 
(\ref{A4PChPTphys}) factorizes in Eq.~(\ref{AphiChPT}). 

Integrating the invariant mass distribution for the $\phi\rightarrow\pi^0\pi^0\gamma$ decay
over the whole physical region one obtains 
$\Gamma(\phi\rightarrow\pi^0\pi^0\gamma)_{\chi}=219$ eV and 
\begin{equation}
\label{BRChPT}
B(\phi\rightarrow\pi^0\pi^0\gamma)_\chi=0.49\times 10^{-4}\ .
\end{equation}
These results improve the prediction for this process given in 
Ref.~\cite{Bramon:1992ki} where $SU(3)$-breaking effects were ignored\footnote{
In the good $SU(3)$ limit, {\it i.e.}~with the chiral amplitude
${\cal A}(K^+K^-\rightarrow\pi^0\pi^0)_\chi=s/4f_\pi f_K\rightarrow s/4f_\pi^2$
and $g_s\rightarrow g\simeq 4.2$ for the $\rho \pi\pi$ strong coupling constant used in
Ref.~\protect\cite{Bramon:1992ki}, one obtains  
$B(\phi\rightarrow\pi^0\pi^0\gamma)_{\chi}=0.54\times 10^{-4}$.}.

\section{Scalar meson exchange in $\phi\rightarrow\pi^0\pi^0\gamma$}
\label{sectLsigmaM}
We now turn to the contributions coming from scalar resonance exchange.
From a ChPT perspective their effects are encoded in the low energy constants of the 
higher order pieces of the ChPT Lagrangian.
But the effects of the $f_0(980)$
---and its lower mass partner $\sigma(500)$ meson, if confirmed---
should manifest in the $\phi\rightarrow\pi^0\pi^0\gamma$ decays not as a constant term
but rather through more complex resonant amplitudes. 
In this section we propose an amplitude which not only obeys the ChPT dictates  
in the lowest part of the $\pi^0\pi^0$ spectrum, as it must,  
but also generates the scalar meson effects for the higher part of the spectrum
where the resonant poles should dominate.

The Linear Sigma Model (L$\sigma$M) \cite{Levy:1967,Gasiorowicz:1969kn,Schechter:1971}, 
which we take as a first guidance probably requiring future refinements,  
will be shown to be particularly appropriate for our purposes. 
It is a well-defined $U(3)\times U(3)$ chiral model which incorporates {\it ab initio}
both the nonet of pseudoscalar mesons together with its chiral partner, 
the scalar mesons nonet.
In this context, the $V\rightarrow P^0P^0\gamma$ decays proceed through a loop of 
charged pseudoscalar mesons emitted by the initial vector. 
Because of the additional emission of a photon, charged pseudoscalar pairs with 
the initial $J^{PC}=1^{--}$ quantum numbers can rescatter into $J^{PC}=0^{++}$ pairs of
charged or neutral pseudoscalars.  
For the $\phi\rightarrow\pi^0\pi^0\gamma$ decay the contributions from charged pion
loops are again negligible compared to those from kaon loops and do not need to be
considered for the same reasons discussed in the previous section.   
The elusive $\sigma(500)$ and the clearly established $f_0(980)$ scalar resonances
are then expected, at least in principle, to play the central role in this
$K^+K^-\rightarrow\pi^0\pi^0$ rescattering process. Their contributions will be conveniently
parametrized in terms of L$\sigma$M amplitudes compatible with ChPT for low dipion invariant
masses. 
 
The $K^+ K^- \rightarrow\pi^0\pi^0$ amplitude in the L$\sigma$M turns out to be  
\begin{equation}
\label{A4PLsigmaM}
\begin{array}{rcl}
{\cal A}(K^+K^- &\!\!\!\! \rightarrow &\!\!\!\! \pi^0\pi^0)_{\mbox{\scriptsize L$\sigma$M}}=
g_{K^+ K^-\pi^0\pi^0}\\[1ex]
&\!\!\!\! - &\!\!\!\!
   \frac{g_{\sigma K^+K^-}g_{\sigma\pi^0\pi^0}}{s-m^2_\sigma}
  -\frac{g_{f_0 K^+K^-}g_{f_0\pi^0\pi^0}}{s-m^2_{f_0}}
  -\frac{g^2_{\kappa^\mp K^\pm\pi^0}}{t-m^2_\kappa} 
  -\frac{g^2_{\kappa^\mp K^\pm\pi^0}}{u-m^2_\kappa}\ ,
\end{array}
\end{equation}
where the various coupling constants are fixed within the model and can be expressed in terms
of $f_{\pi}$, $f_{K}$, 
the masses of the pseudoscalar and scalar mesons involved in the process,
and the scalar meson mixing angle in the flavour basis 
$\phi_S$ \cite{Napsuciale:1998ip,Tornqvist:1999tn}.  
In particular, the $g_{K^+ K^-\pi^0\pi^0}$ coupling accounting for the constant
four-pseudoscalar amplitude can be expressed in a more convenient form by imposing that
the amplitude ${\cal A}(K^+K^-\rightarrow\pi^0\pi^0)_{\mbox{\scriptsize L$\sigma$M}}$
vanishes in the soft-pion limit (either $p\rightarrow 0$ or $p'\rightarrow 0$).  
Then, the amplitude (\ref{A4PLsigmaM}) can be rewritten as the sum of three terms
each one depending only on $s$, $t$ and $u$: 
\begin{equation}
\label{A4PLsigmaMphys}
\begin{array}{l}
{\cal A}(K^+K^-\rightarrow\pi^0\pi^0)_{\mbox{\scriptsize L$\sigma$M}}\equiv
{\cal A}_{\mbox{\scriptsize L$\sigma$M}}^s (s)+
{\cal A}_{\mbox{\scriptsize L$\sigma$M}}^t (t)+
{\cal A}_{\mbox{\scriptsize L$\sigma$M}}^u (u)\\[1ex]
\ \ \ \ =
\frac{s-m^2_{\pi}}{2f_\pi f_K}\left[
\frac{m^2_K-m^2_{\sigma}}{D_{\sigma}(s)}{\rm c}\phi_S({\rm c}\phi_S -\sqrt{2}{\rm s}\phi_S)+
\frac{m^2_K-m^2_{f_0}}{D_{f_0}(s)}{\rm s}\phi_S({\rm s}\phi_S +\sqrt{2}{\rm c}\phi_S)
\right]\\[2ex]
\ \ \ \ +
\frac{t-m^2_K}{4f_\pi f_K}\frac{m^2_\pi-m^2_{\kappa}}{D_{\kappa}(t)}+
\frac{u-m^2_K}{4f_\pi f_K}\frac{m^2_\pi-m^2_{\kappa}}{D_{\kappa}(u)}\ ,
\end{array}
\end{equation}
where $D_{S}(s)=s-m^2_S +i\,m_S\Gamma_S$ are the $S=\sigma, f_0$ propagators   
---similar expressions hold for $D_{\kappa}(t,u)$---  
and $({\rm c}\phi_S, {\rm s}\phi_S)\equiv (\cos\phi_S, \sin\phi_S)$. 
Notice that we start using simple Breit-Wigner expressions for the propagators in spite of 
the well known difficulties associated to threshold effects for the $f_0$. Moreover, 
the amplitude (\ref{A4PLsigmaMphys}) is not strictly unitary.
However, we will not discuss this issue in detail but refer the reader to 
Ref.~\cite{Ishida:1997wn} where a careful study of full unitarity in $K\pi$ scattering
in the framework of the L$\sigma$M can be found.
We should stress that the main concern of our analysis is to explicitly
incorporate the scalar resonances and that, for this specific purpose,
the tree level (and not fully unitary) amplitude (\ref{A4PLsigmaMphys}) 
with simple expressions for the scalar propagators will be shown to be a 
reasonable starting point.
A more sophisticated treatment of the $f_0$ propagator 
\cite{Achasov:1987ts,Flatte:1972rz,Escribano:2002iv} accounting for the $K \bar{K}$
threshold effects ($m_{f_0} \simeq 2m_K$) will be discussed later on. 

A few remarks on the four-pseudoscalar amplitudes in 
Eqs.~(\ref{A4PLsigmaM},\ref{A4PLsigmaMphys}) and on their comparison with the
ChPT amplitude in Eq.~(\ref{A4PChPTphys}) are of interest:
\begin{itemize}
\item[i)]
for $m_S\rightarrow\infty$ ($S=\sigma, f_0, \kappa$),
the L$\sigma$M amplitude (\ref{A4PLsigmaMphys}) reduces to 
\[
\frac{s-m^2_\pi}{2f_\pi f_K}+\frac{t+u-2m^2_K}{4f_\pi f_K}\ ,
\]
which would coincide with the ChPT amplitude (\ref{A4PChPTphys})
if the on-shell condition $s+t+u=2m^2_\pi+2m^2_K$ could be invoked. 
As shown in Eq.~(\ref{A4PLsigmaM}),
${\cal A}(K^+K^-\rightarrow\pi^0\pi^0)_{\mbox{\scriptsize L$\sigma$M}}$
consists of a constant four-pseudoscalar vertex plus three terms 
whose $s,t,u$ dependence is generated by the scalar propagators $D_S(s,t,u)$. 
Combining each one of these three terms with its corresponding part from the first, 
four-pseudoscalar vertex leads immediately to Eq.~(\ref{A4PLsigmaMphys}).
For $m_S\rightarrow\infty$, this amplitude is linear in the $s,t,u$ variables,
mimics perfectly the effects of the derivative and massive terms in the Lagrangian
(\ref{ChPTlag}) and leads to the ChPT amplitude (\ref{A4PChPTphys}) once
$s+t+u=2m^2_K+2m^2_\pi$ is used.  
This corresponds to the aforementioned complementarity between ChPT and the L$\sigma$M,
thus making the whole analysis quite reliable. 

\item[ii)]
however, the relation $s+t+u=2m^2_K+2m^2_\pi$ cannot be invoked when plugging our
L$\sigma$M amplitude (\ref{A4PLsigmaMphys}) into the loop of {\it virtual} kaons leading to
the $\phi\rightarrow\pi^0\pi^0\gamma$ decay amplitude.
This is in sharp contrast with the processes studied in
Refs.~\cite{Bramon:2000vu,Escribano:2000fs,Bramon:2001un},
governed by $t,u$-independent L$\sigma$M amplitudes, and thus requires a new treatment for our
case. To this aim, one has to distinguish between the ${\cal A}^s_{\mbox{\scriptsize
L$\sigma$M}}$ term in Eq.~(\ref{A4PLsigmaMphys}) and the    
${\cal A}^t_{\mbox{\scriptsize L$\sigma$M}}+{\cal A}^u_{\mbox{\scriptsize L$\sigma$M}}$ terms 
of the same amplitude. 
The former, being only $s$-dependent, can be directly introduced in Eq.~(\ref{AphiChPT})
instead of the ${\cal A}(K^+ K^-\rightarrow\pi^0\pi^0)_\chi$ amplitude. Fortunately, this term
contains most of the relevant dynamics due to the scalar resonances ---the $f_0$(980) and,
eventually, the $\sigma$(500) poles--- for the process $\phi\rightarrow\pi^0\pi^0\gamma$ where
the dipion mass spectrum covers the  range $4 m_\pi^2 \le m^2_{\pi\pi} \le m^2_\phi$.  
The  remaining two terms  
${\cal A}^t_{\mbox{\scriptsize L$\sigma$M}}+{\cal A}^u_{\mbox{\scriptsize L$\sigma$M}}$, 
being $t,u$-dependent, cannot be simply plugged and evaluated into the
loop integral\footnote{A more involved four-propagator loop integral would be required
to compute the contributions of the $t,u$-channel in a precise way.}
but their contribution to the $\phi\rightarrow\pi^0\pi^0\gamma$ amplitude
can be identified and reasonably estimated.
Indeed, the $\kappa$ contribution to the
${\cal A}(K^+K^-\rightarrow\pi^0\pi^0)_{\mbox{\scriptsize L$\sigma$M}}$
amplitude can be fixed by subtracting from the chiral-loop amplitude
in Eq.~(\ref{A4PChPTphys}) the contributions from 
the $\sigma$ and $f_0$ terms in Eq.~(\ref{A4PLsigmaMphys}) taking $m_{\sigma,
f_0}\rightarrow\infty$. The resulting expression corresponds to the desired $\kappa$
contribution in the $m_\kappa\rightarrow\infty$ limit.
The difference between this $\kappa$ contribution for $m_\kappa\rightarrow\infty$ and the same
contribution for a $\kappa$ of finite mass can be considered as negligible
due to the high mass of the $\kappa$ and the lack of scalar poles in 
the $t$- and $u$-channel of the $\phi\rightarrow\pi^0\pi^0\gamma$ decay process\footnote{
The $\kappa$ has been identified in the literature as a broad resonance with a mass around 900 MeV
(see for example Ref.~\protect\cite{Napsuciale:1998ip}) or with the $K_0^\ast(1430)$ scalar
resonance as in Ref.~\protect\cite{Tornqvist:1999tn}.
A model-independent analysis based on the analytic continuation of $\pi K$ scattering finds that
there is room for a $K^*_0(1430)$ scalar state but not for a $\kappa(900)$ one
\protect\cite{Cherry:2000ut}.}. 
In other words, while the $K^+K^-\rightarrow\pi^0\pi^0$ rescattering amplitude evaluated at
lowest order in ChPT is necessarily a poor approximation to the (pole dominated) $s$-channel
dynamics in $\phi\rightarrow\pi^0\pi^0\gamma$ decays, the same lowest order amplitude can
be taken as a reasonable estimate for the $t$- and $u$-channel contributions where the  
$\kappa$ pole position cannot be kinematically approached.  
Therefore, an improved expression for the chiral-loop amplitude in Eq.~(\ref{A4PChPTphys})
is the following:  
\begin{equation}
\label{AphiChPT+LsigmaM}
\begin{array}{l}
{\cal A}(K^+K^-\rightarrow\pi^0\pi^0)_{\mbox{\scriptsize L$\sigma$M}}=
\frac{m^2_{\pi}-s/2}{2f_\pi f_K}+\frac{s-m^2_{\pi}}{2f_\pi f_K}\\[2ex]
\ \ \times\left[
\frac{m^2_K-m^2_{\sigma}}{D_{\sigma}(s)}{\rm c}\phi_S({\rm c}\phi_S -\sqrt{2}{\rm s}\phi_S)+
\frac{m^2_K-m^2_{f_0}}{D_{f_0}(s)}{\rm s}\phi_S({\rm s}\phi_S +\sqrt{2}{\rm c}\phi_S)
\right]\ .
\end{array}
\end{equation}
As desired, it has no $t,u$ dependence and can be thus plugged in Eq.~(\ref{AphiChPT})
instead of ${\cal A}(K^+K^-\rightarrow\pi^0\pi^0)_\chi$. One then obtains the $s$-dependent
amplitude  
\begin{equation}
\label{AphiLsigmaM}
{\cal A}(\phi\rightarrow\pi^0\pi^0\gamma)_{\mbox{\scriptsize L$\sigma$M}}= 
\frac{eg_{s}}{2\pi^2 m^2_{ K^+}}\,\{a\}\,L(s)\times
{\cal A}(K^+ K^-\rightarrow\pi^0\pi^0)_{\mbox{\scriptsize L$\sigma$M}} \ ,
\end{equation}
which will be used from now on.  
Notice that for large scalar masses one recovers Eq.~(\ref{AphiChPT}), 
that the $f_0$(980) (and, eventually, $\sigma$(500)) $s$-channel poles do now appear, 
and that the remaining term, $(m^2_\pi-s/2)/2f_\pi f_K$, accounts for $\kappa$ exchange
effects in the $t,u$-channel where, as stated before, no poles are approached and has thus
been approximated by the corresponding part of the simple chiral-loop amplitude. 

\item[iii)]
the large widths of the scalar resonances break chiral symmetry if they are naively
introduced in Eq.~(\ref{A4PLsigmaM}), an effect already noticed in Ref.~\cite{Achasov:1994iu}.
Accordingly, we introduce the $\sigma(500)$ and $f_0(980)$ widths in the propagators
only {\it after} chiral cancellation of constant terms in the amplitude.
In this way the pseudo-Goldstone nature of pions is preserved.

\item[iv)]
the $\pi^0\pi^0$ invariant mass spectrum for the $\phi\rightarrow\pi^0\pi^0\gamma$ decay
covers the region where the presence of the $f_0(980)$ (and the $\sigma(500)$) meson(s)
should manifest.
Because of the presence of the corresponding propagators in 
Eq.~(\ref{AphiChPT+LsigmaM})
---closely linked to the ChPT amplitude and thus expected to account for the lowest 
   part of the $\pi^0\pi^0$ spectrum--- one 
should also be able to reproduce the effects of the $f_0$ (and the $\sigma(500)$) pole(s)
at higher $\pi^0\pi^0$ invariant mass values.
\end{itemize}

\section{Vector meson exchange in $\phi\rightarrow\pi^0\pi^0\gamma$}
\label{sectVMD}
In addition to the L$\sigma$M contributions, which can be viewed as an
improved version of the chiral-loop predictions,  the analysis should be
extended to include vector meson exchange in the $t$- and $u$-channel. 
These VMD contributions were already considered in Ref.~\cite{Bramon:1992kr}. 
In this framework, $\phi\rightarrow\pi^0\pi^0\gamma$ proceeds through $\omega$-$\phi$
mixing followed by the exchange of intermediate $\rho^0$ mesons in the direct or crossed 
channel of the  
$\phi\rightarrow\pi^0\rho^0\rightarrow\pi^0\pi^0\gamma$ decay chain.

In order to describe these vector meson contributions we use the $SU(3)$-symmetric Lagrangians
\begin{equation}
\label{VMDLag}
\begin{array}{rcl}
{\cal L}_{\rm VVP}&=&\frac{G}{\sqrt{2}}\epsilon^{\mu\nu\alpha\beta}
                     \langle\partial_\mu V_\nu\partial_\alpha V_\beta P\rangle\ ,\\[2ex]
{\cal L}_{{\rm V}\gamma}&=& -4f^2 e g A_\mu\langle Q V^\mu\rangle\ ,
\end{array}
\end{equation}
where $G=\frac{3g^2}{4\pi^2 f}$ is the $\omega\rho\pi$ coupling constant and
$|g|\simeq 4.0$ as follows from various $\rho$ and $\omega$ decay data
\cite{daphne95:bramon,Groom:2000in}. The VMD amplitude for
$\phi(q^\ast,\epsilon^\ast)\rightarrow\pi^0(p)\pi^0(p^\prime)\gamma(q,\epsilon)$
is then found to be  
\begin{equation}
\label{AphiVMD}
\textstyle
{\cal A}(\phi\rightarrow\pi^0\pi^0\gamma)_{\rm VMD}=
\frac{\epsilon}{3}\frac{G^2 e}{\sqrt{2}g}
\left(\frac{P^2\{a\}+\{b(P)\}}{M^2_\rho-P^2-i M_\rho\Gamma_\rho}+
      \frac{{P^\prime}^2\{a\}+\{b({P^\prime})\}}
           {M^2_\rho-{P^\prime}^2-i M_\rho\Gamma_\rho}\right)\ ,
\end{equation}
with $\{a\}$ as in Eq.~(\ref{AphiChPT}) and a new amplitude 
\begin{equation}
\label{b}
\begin{array}{rcl}
\{b(P)\}&=& -(\epsilon^\ast\cdot\epsilon)\,(q^\ast\cdot P)\,(q\cdot P)
            -(\epsilon^\ast\cdot P)\,(\epsilon\cdot P)\,(q^\ast\cdot q)\\[1ex]
        & & +(\epsilon^\ast\cdot q)\,(\epsilon\cdot P)\,(q^\ast\cdot P)
            +(\epsilon\cdot q^\ast)\,(\epsilon^\ast\cdot P)\,(q\cdot P)\ ,
\end{array}
\end{equation}
where $P=p+q$ and $P^\prime=p^\prime+q$ are the momenta of the intermediate
$\rho^0$ meson in the $t$- and $u$-channel, respectively.
From this VMD amplitude one easily obtains 
$\Gamma(\phi\rightarrow\pi^0\pi^0\gamma)_{\rm VMD}=37$ eV and 
\begin{equation}
\label{BRVMD}
B(\phi\rightarrow\pi^0\pi^0\gamma)_{\rm VMD}=8.3\times 10^{-6}\ ,
\end{equation}
in agreement with the results in Ref.~\cite{Bramon:1992kr,Achasov:2001cj}
once the same numerical inputs are used.
Further details on these contributions are given in the following section. 

\section{Final results}
Our final results for ${\cal A}(\phi\rightarrow\pi^0\pi^0\gamma)$  
are thus the sum of the VMD contribution in Eq.~(\ref{AphiVMD}) plus the L$\sigma$M  
contribution containing the scalar resonance effects in Eq.~(\ref{AphiLsigmaM}). 
The final $\pi^0\pi^0$ invariant mass distribution for the process is\footnote{
In terms of the photon energy, $E_\gamma=(m^2_\phi-m^2_{\pi^0\pi^0})/(2m_\phi)$,
the photonic spectrum is written as
$d\Gamma/dE_\gamma=(m_\phi/m_{\pi^0\pi^0})\times d\Gamma/dm_{\pi^0\pi^0}$.}
\begin{equation}
\label{dGdmtotal}
\frac{d\Gamma(\phi\rightarrow\pi^0\pi^0\gamma)}{dm_{\pi^0\pi^0}}=
\frac{d\Gamma_{\mbox{\scriptsize L$\sigma$M}}}{dm_{\pi^0\pi^0}}+
\frac{d\Gamma_{\mbox{\scriptsize VMD}}}{dm_{\pi^0\pi^0}}+
\frac{d\Gamma_{\mbox{\scriptsize int}}}{dm_{\pi^0\pi^0}}\ ,
\end{equation}
where the L$\sigma$M term (the scalar or \emph{signal} contribution) is
\begin{equation}
\label{dGLsigmaM}
\begin{array}{rl}
\frac{d\Gamma_{\mbox{\tiny L$\sigma$M}}}{dm_{\pi^0\pi^0}}&=\ 
\frac{1}{2}\frac{\alpha}{192\pi^5}\frac{g_{s}^2}{4\pi}\frac{m^4_\phi}{m^4_{K^+}}
\frac{m_{\pi^0\pi^0}}{m_\phi}\left(1-\frac{m^2_{\pi^0\pi^0}}{m^2_\phi}\right)^3
\sqrt{1-\frac{4m^2_{\pi^0}}{m^2_{\pi^0\pi^0}}}\\[2ex]
&\times\ |L(m^2_{\pi^0\pi^0})|^2
|{\cal A}(K^+K^-\rightarrow\pi^0\pi^0)_{\mbox{\scriptsize L$\sigma$M}}|^2\ ,
\end{array}
\end{equation}
with the four-pseudoscalar amplitude taken from Eq.~(\ref{AphiChPT+LsigmaM}). 
The VMD term (the vector or \emph{background} contribution)
and the contribution resulting from the interference of both amplitudes
are given by
\begin{equation}
\label{dGdmVMDint}
\begin{array}{rl}
\frac{d\Gamma_{\mbox{\tiny [VMD,int]}}}{dm_{\pi^0\pi^0}}&=\
\frac{1}{2}\frac{1}{256\pi^3}
\frac{m_{\pi^0\pi^0}}{m_\phi}\left(1-\frac{m^2_{\pi^0\pi^0}}{m^2_\phi}\right)
\sqrt{1-\frac{4m^2_{\pi^0}}{m^2_{\pi^0\pi^0}}}\\[2ex]
&\times\ {\displaystyle\int_{-1}^{1}}dx A_{\rm [VMD,int]}(m_{\pi^0\pi^0},x)\ ,
\end{array}
\end{equation}
where one explicitly has  
\begin{equation}
\label{AVMD2}
\begin{array}{l}
A_{\rm VMD}(m_{\pi^0\pi^0},x)\equiv 
\frac{1}{3}\sum_{\rm pol}|{\cal A}_{\rm VMD}|^2=
\frac{1}{3}\left(\frac{\epsilon}{3}\frac{G^2 e}{\sqrt{2}g}\right)^2\\[2ex]
\ \times
\left[\frac{1}{8}\left(
m^8_{\pi^0}-2m^2_\phi m^6_{\pi^0}+4\tilde m^4_{\rho}m^4_{\pi^0}+m^4_\phi m^4_{\pi^0}
-2\tilde m^2_{\rho}\tilde m^{\ast 2}_{\rho} m^4_{\pi^0}-4\tilde m^6_{\rho}m^2_{\pi^0}
\right.\right.
\\[2ex]
\ \ \ \ 
-2\tilde m^2_{\rho}m^4_\phi m^2_{\pi^0}+4\tilde m^4_{\rho}m^2_\phi m^2_{\pi^0}
-4\tilde m^4_{\rho}\tilde m^{\ast 2}_{\rho} m^2_{\pi^0}
+2\tilde m^2_{\rho}m^2_\phi\tilde m^{\ast 2}_{\rho} m^2_{\pi^0}+\tilde m^8_{\rho}
\\[2ex]
\ \ \ \;
\left.
+\tilde m^4_{\rho}m^4_\phi+2\tilde m^4_{\rho}\tilde m^{\ast 4}_{\rho}
-2\tilde m^6_{\rho}m^2_\phi+2\tilde m^6_{\rho}\tilde m^{\ast 2}_{\rho}
-2\tilde m^4_{\rho}m^2_\phi\tilde m^{\ast 2}_{\rho}\right)
\times\frac{1}{|D_{\rho}(\tilde m^2_{\rho})|^2}
\\[2ex]
\
+\frac{1}{16}\left(
m^8_{\pi^0}-\tilde m^2_{\rho}m^2_\phi m^4_{\pi^0}+
2\tilde m^2_{\rho}\tilde m^{\ast 2}_{\rho}m^4_{\pi^0}-
m^2_\phi\tilde m^{\ast 2}_{\rho}m^4_{\pi^0}-
4\tilde m^2_{\rho}\tilde m^{\ast 4}_{\rho}m^2_{\pi^0}\right.
\\[2ex]
\ \ \ \
-4\tilde m^4_{\rho}\tilde m^{\ast 2}_{\rho}m^2_{\pi^0}+
4\tilde m^2_{\rho}m^2_\phi\tilde m^{\ast 2}_{\rho}m^2_{\pi^0}+
\tilde m^2_{\rho}\tilde m^{\ast 6}_{\rho}+3\tilde m^4_{\rho}\tilde m^{\ast 4}_{\rho}-
\tilde m^2_{\rho}m^2_\phi\tilde m^{\ast 4}_{\rho}
\\[2ex]
\ \ \ \; 
\left.\left.+\tilde m^6_{\rho}\tilde m^{\ast 2}_{\rho}-
\tilde m^4_{\rho}m^2_\phi\tilde m^{\ast 2}_{\rho}\right)
\times
2\mbox{Re}\left(\frac{1}{D_{\rho}(\tilde m^2_{\rho})
                         D_{\rho}^\ast(\tilde m^{\ast 2}_{\rho})}\right)
+\left(\tilde m^2_{\rho}\leftrightarrow \tilde m^{\ast 2}_{\rho}\right)\right]\ ,\\[2ex]
\end{array}
\end{equation}
and 
\begin{equation}
\label{Aint}
\begin{array}{l}
A_{\rm int}(m_{\pi^0\pi^0},x)\equiv
\frac{2}{3}\mbox{Re}\sum_{\rm pol}{\cal A}_{\mbox{\scriptsize L$\sigma$M}}{\cal A}_{\rm VMD}^\ast=
\frac{1}{3}\left(\frac{eg_{s}}{2\pi^2 m^2_{K^+}}\right)
\left(\frac{\epsilon}{3}\frac{G^2 e}{\sqrt{2}g}\right)\\[2ex]
\ \times\,2\mbox{Re}\left\{L(m^2_{\pi^0\pi^0})
{\cal A}(K^+ K^-\rightarrow\pi^0\pi^0)_{\mbox{\scriptsize L$\sigma$M}}\right.\\[2ex]
\ \times
\left.\frac{1}{4}\left[
\frac{\tilde m^2_{\rho}(\tilde m^2_{\rho}+\tilde m^{\ast 2}_{\rho}-2m^2_{\pi^0})^2-
m^2_\phi(\tilde m^2_{\rho}-m^2_{\pi^0})^2}{D_{\rho}^\ast(\tilde m^2_{\rho})}
+\left(\tilde m^2_{\rho}\leftrightarrow \tilde m^{\ast 2}_{\rho}\right)
\right]\right\}\ ,\\[2ex]
\end{array}
\end{equation}
with
\begin{equation}
\label{mrhotilde}
\begin{array}{rcl}
P^2\equiv\tilde m^2_{\rho}&=&
m^2_{\pi^0}+\frac{m^2_\phi-m^2_{\pi^0\pi^0}}{2}
\left(1-x\sqrt{1-\frac{4m^2_{\pi^0}}{m^2_{\pi^0\pi^0}}}\right)\ ,\\[2ex]
{P^\prime}^2\equiv\tilde m^{\ast 2}_{\rho}&=&
m^2_{\pi^0}+\frac{m^2_\phi-m^2_{\pi^0\pi^0}}{2}
\left(1+x\sqrt{1-\frac{4m^2_{\pi^0}}{m^2_{\pi^0\pi^0}}}\right)\\[2ex]
&=&2m_{\pi^0}^2+m^2_\phi-m^2_{\pi^0\pi^0}-\tilde m^2_{\rho}
\ .\\[2ex]
\end{array}
\end{equation}

\begin{figure}[t]
\centerline{\includegraphics[width=0.85\textwidth]{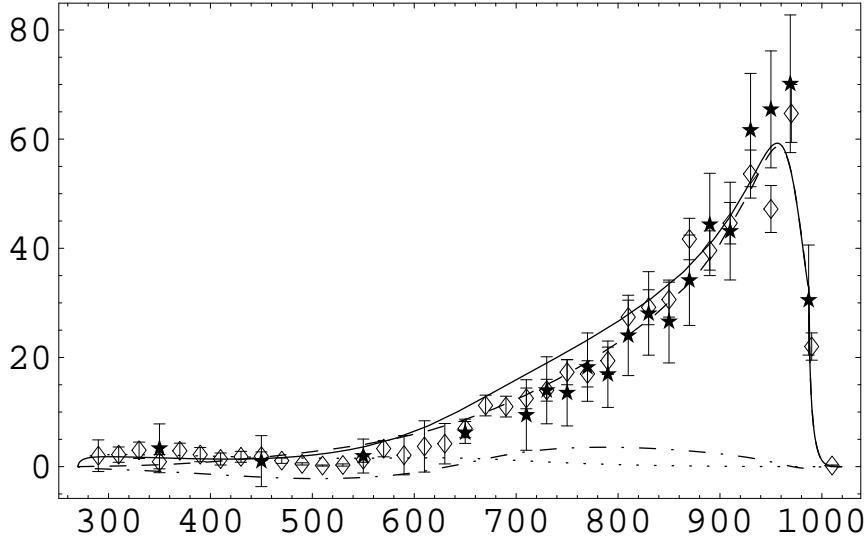}} 
\caption{\small
$dB(\phi\rightarrow\pi^0\pi^0\gamma)/dm_{\pi^0\pi^0} \times 10^8$ (in units of MeV$^{-1}$)   
as a function of the dipion invariant mass $m_{\pi^0\pi^0}$ (in MeV).
The dashed, dotted and dot-dashed lines correspond to the separate contributions 
from L$\sigma$M, VMD and their interference, respectively. 
The solid line is the total result. 
We have taken $m_\sigma=478$ MeV, $\Gamma_\sigma=324$ MeV,
$m_{f_{0}}=980$ MeV, $\Gamma_{f_{0}}=70$ MeV and $\phi_{S}=-6^\circ$.
Simple Breit-Wigner formulae have been used for the $\sigma$ and $f_0$ propagators.
Experimental data $(\star)$ are taken from Ref.~\protect\cite{Achasov:2000ym}
and $(\diamond)$ from Ref.~\protect\cite{Aloisio:2002bt}.} 
\label{dBdmallcontribBW}
\end{figure}

The separate contributions from L$\sigma$M, VMD and their interference,  
as well as the total result are shown in Fig.~\ref{dBdmallcontribBW}.
In this figure, whose main purpose is to illustrate the possibilities of our approach,
we use simple Breit-Wigner formulae for the $\sigma$ and the $f_0$ propagators. 
We take $m_{\sigma}=478$ MeV and $\Gamma_{\sigma}=324$ MeV, 
which are the central values of the measurements quoted in \cite{Aitala:2001xu},    
$m_{f_{0}}=980$ MeV and $\Gamma_{f_{0}}=70$ MeV, from Ref.~\cite{Groom:2000in},   
and $\phi_{S}=-6^\circ$ for the scalar mixing angle.
The global agreement with the data is rather good although the model will require
further improvements.
As expected, $f_{0}(980)$ scalar meson exchange contributes decisively to achieve this 
agreement. Indeed, for the integrated decay width one now obtains 
$\Gamma(\phi\rightarrow\pi^0\pi^0\gamma)_{\mbox{\scriptsize L$\sigma$M+VMD}}=530$ eV 
and for the branching ratio 
\begin{equation} 
\label{BRLsigmaM+VMD} 
B(\phi\rightarrow\pi^0\pi^0\gamma)_{\mbox{\scriptsize L$\sigma$M+VMD}}= 
1.19\times 10^{-4}\ , 
\end{equation} 
quite in line with the experimental results quoted in Eqs.~(\ref{VEPP}) and (\ref{KLOE}). 
Without the VMD contribution, our result decreases only by some 10\%,    
$B(\phi\rightarrow\pi^0\pi^0\gamma)_{\mbox{\scriptsize L$\sigma$M}}= 
1.07\times 10^{-4}\ $, but it still remains well above the value quoted in
Eq.~(\ref{BRChPT}),  
$B(\phi\rightarrow\pi^0\pi^0\gamma)_\chi=0.49\times 10^{-4}$,
which did not contain the crucial scalar contributions.   
\begin{figure}[t]
\centerline{\includegraphics[width=0.5\textwidth]{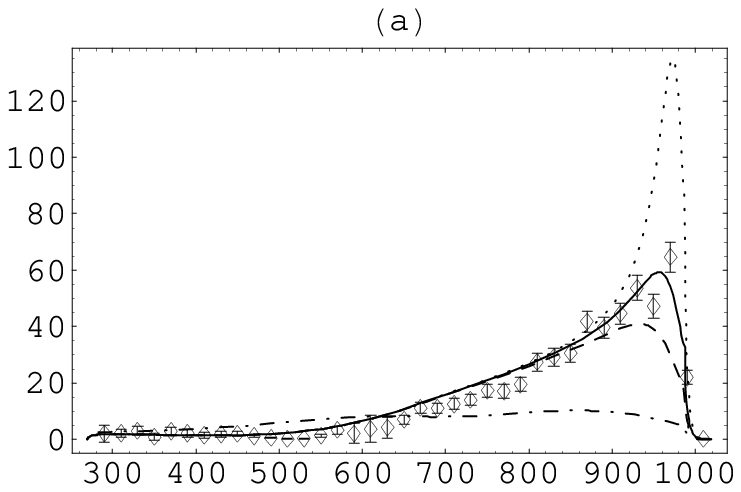}
            \includegraphics[width=0.5\textwidth]{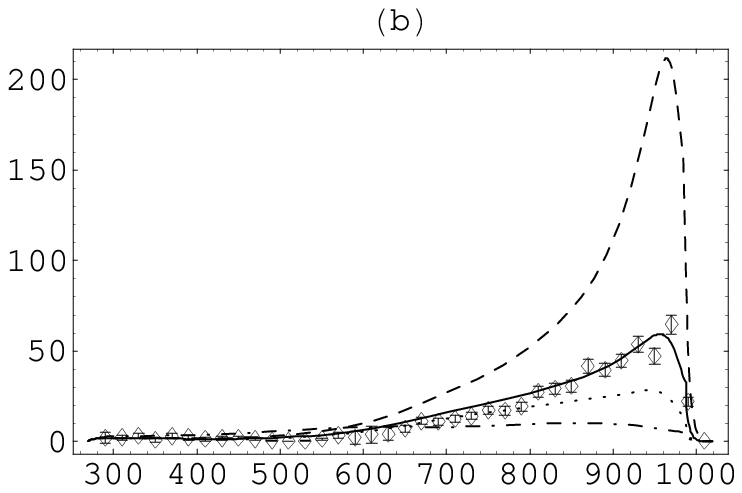}}  
\caption{\small 
$dB(\phi\rightarrow\pi^0\pi^0\gamma)/dm_{\pi^0\pi^0} \times 10^8$ (in units of MeV$^{-1}$)   
as a function of the dipion invariant mass $m_{\pi^0\pi^0}$ (in MeV). 
Simple Breit-Wigner formulae have been used for the $\sigma$ and $f_0$ propagators. 
(a) predictions for a fixed scalar mixing angle
$\phi_{S}=-6^\circ$ and
$\Gamma_{f_{0}}=40$ MeV (dotted line),
$\Gamma_{f_{0}}=70$ MeV (solid line) or 
$\Gamma_{f_{0}}=100$ MeV (dashed line).
(b) predictions  for a fixed $f_{0}$ decay width
$\Gamma_{f_{0}}=70$ MeV and
$\phi_{S}=-3^\circ$ (dotted line),
$\phi_{S}=-6^\circ$ (solid line) or 
$\phi_{S}=-14^\circ$ (dashed line).
The prediction including chiral loops and VMD corrections is also shown for comparison
(dot-dashed line). Experimental data are taken from Ref.~\protect\cite{Aloisio:2002bt}.} 
\label{dBdmGf0phiS} 
\end{figure}
In order to show the sensitivity of our treatment on the $f_{0}$ decay width and on
the scalar mixing angle we have plotted in Fig.~\ref{dBdmGf0phiS} our predictions for the
$\pi^0\pi^0$ invariant mass distribution in $\phi\rightarrow\pi^0\pi^0\gamma$ for
various values of $\Gamma_{f_{0}}$ and $\phi_{S}$.
The prediction including chiral-loops (without scalar poles) and VMD corrections is also
shown for comparison. 
The invariant mass spectrum and the branching ratio are very sensitive to $\Gamma_{f_{0}}$
and, even more, to $\phi_{S}$. 
Present data on $\phi\rightarrow\pi^0\pi^0\gamma$ 
\cite{Achasov:2000ym,Akhmetshin:1999di,Aloisio:2002bt} favour $\phi_{S} \simeq -6^\circ$
---which is in the expected range $-14^\circ \leq \phi_{S} \leq -3^\circ$
   \cite{Napsuciale:1998ip,Tornqvist:1999tn}---
and $\Gamma_{f_0} \simeq 70$ MeV for the total width of the $f_0$ Breit-Wigner.
This latter value is compatible with the PDG estimates, 
$\Gamma_{f_0}\simeq 40$--$100$ MeV, and somewhat above other recent values obtained with
usual Breit-Wigner expressions:
$\Gamma_{f_{0}}=56\pm 20$ MeV \cite{Akhmetshin:1999di} or $44\pm 2\pm 2$ MeV \cite{Aitala:2000xt}. 
Notice that in our approach $\phi_{S} \simeq -6^\circ$ implies
$\Gamma (f_{0}\rightarrow\pi\pi)\simeq  35$ MeV accounting for some 50\% of the total $f_0$-width.
Other decay channels, such as $f_0\rightarrow K\bar{K}$ for which our treatment predicts a
large coupling, should account for the remaining 50\%.
But this requires a two-channel analysis around the $f_0\rightarrow K\bar{K}$ threshold
to which we now turn. 

\begin{figure}[t]
\centerline{\includegraphics[width=0.85\textwidth]{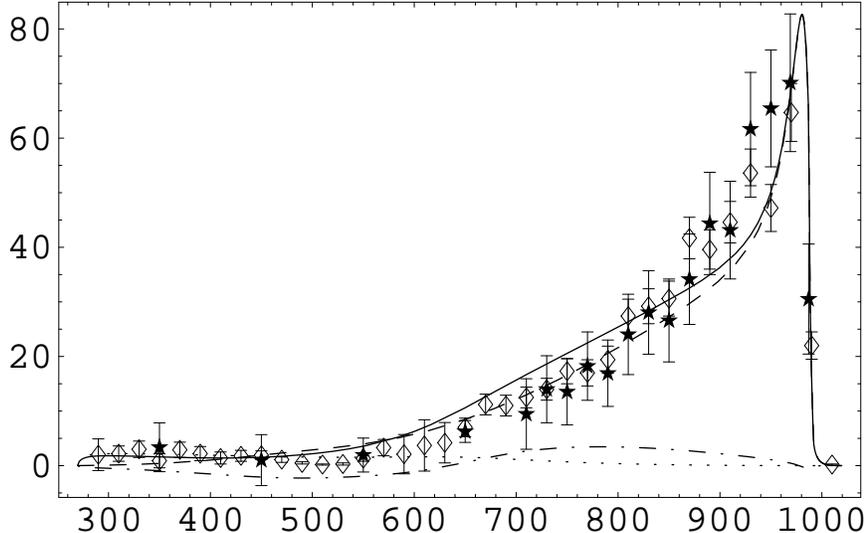}} 
\caption{\small
$dB(\phi\rightarrow\pi^0\pi^0\gamma)/dm_{\pi^0\pi^0} \times 10^8$ (in units of MeV$^{-1}$)   
as a function of the dipion invariant mass $m_{\pi^0\pi^0}$ (in MeV). 
The dashed, dotted and dot-dashed lines correspond to the separate contributions 
from L$\sigma$M, VMD and their interference, respectively. 
The solid line is the total result. 
For the $\sigma$ propagator we use a simple Breit-Wigner with $m_\sigma=478$ MeV and 
$\Gamma_\sigma=324$ MeV.
For the $f_0$ propagator we use the complete one-loop expression 
\protect\cite{Achasov:1987ts,Escribano:2002iv} with $m_{f_{0}}=985$ MeV and
$\phi_{S}=-9^\circ$.
Experimental data $(\star)$ are taken from Ref.~\protect\cite{Achasov:2000ym}
and $(\diamond)$ from Ref.~\protect\cite{Aloisio:2002bt}.}
\label{dBdmallcontribF}
\end{figure}

\begin{figure}[t]
\centerline{\includegraphics[width=0.85\textwidth]{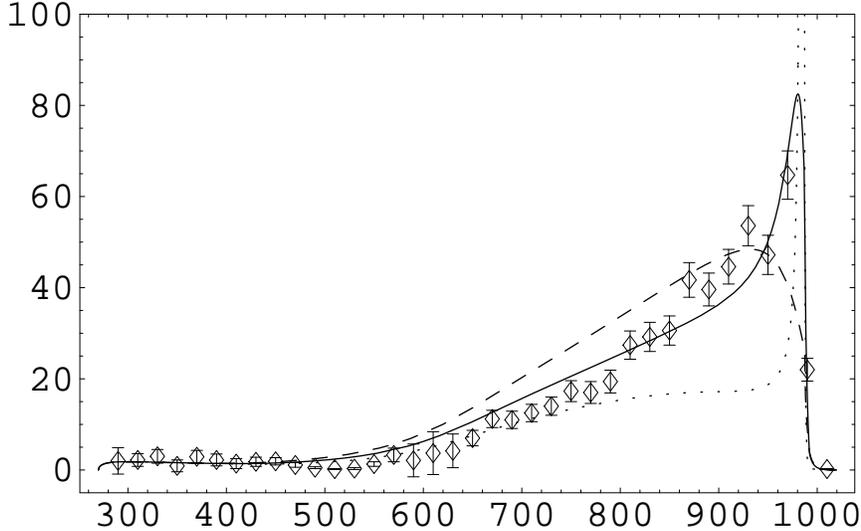}}  
\caption{\small 
$dB(\phi\rightarrow\pi^0\pi^0\gamma)/dm_{\pi^0\pi^0} \times 10^8$ (in units of MeV$^{-1}$)   
as a function of the dipion invariant mass $m_{\pi^0\pi^0}$ (in MeV). 
The complete one-loop expression has been used for the $f_0$ propagator. 
Predictions for a fixed $f_{0}$ mass $m_{f_{0}}=985$ MeV and
$\phi_{S}=-3^\circ$ (dotted line),
$\phi_{S}=-9^\circ$ (solid line) or 
$\phi_{S}=-14^\circ$ (dashed line).
Experimental data are taken from Ref.~\protect\cite{Aloisio:2002bt}.} 
\label{dBdmphiS} 
\end{figure}

The original two-channel treatment of scalar resonances near the
$K\bar{K}$ threshold due to Flatt\'e \cite{Flatte:1972rz} has been improved in 
Refs.~\cite{Achasov:1987ts,Escribano:2002iv}
and is now widely applied in experimental analyses of $V\rightarrow P P\gamma$
decays \cite{Achasov:2000ym,Akhmetshin:1999di,Aloisio:2002bt,Aloisio:2002bs,Achasov:2000ku}.
It amounts to the substitution of the simple expression 
$D_{f_0}(s)=s-m^2_{f_0}+i\,m_{f_0}\Gamma_{f_0}$ by 
a complete one-loop $f_0$ propagator $D_{f_0}(s)=s-m^2_{f_0}-\mbox{Re}\Pi(m^2_{f_0})+\Pi(s)$,
where the term $\Pi(s)-\mbox{Re}\Pi(m^2_{f_0})$ takes into account the finite-width corrections
due to the $\pi\pi$ and $K\bar K$ channels.
Within our L$\sigma$M approach, the $f_0\rightarrow\pi\pi$ and $f_0\rightarrow K\bar{K}$
partial widths can be expressed in terms of $m_{f_0}$, $\phi_{S}$ and accurately measured
pseudoscalar masses and decay constants. 
For $m_{f_0} = 985$ MeV and $\phi_{S} = -9^\circ$ we obtain the various curves shown in 
Fig.~\ref{dBdmallcontribF} which explain quite reasonably the available data. 
Integrating over the whole kinematical region one obtains
$\Gamma (\phi\rightarrow\pi^0\pi^0\gamma) = 518$ eV and 
\begin{equation}
\label{BRLsigmaM+VMDFlatte}
B(\phi\rightarrow\pi^0\pi^0\gamma)=1.16\times 10^{-4}\ .
\end{equation}
The shape of the $\pi\pi$ mass spectrum depends strongly on the value for $\phi_{S}$,
as shown in Fig.~\ref{dBdmphiS}, and follows the same pattern previously shown in
Fig.~\ref{dBdmGf0phiS}. 

\section{Comments and conclusions}
The recent publication by experimental groups working at
VEPP-2M \cite{Achasov:2000ym,Akhmetshin:1999di} and DA$\Phi$NE \cite{Aloisio:2002bt}
of the first data on the $\phi\rightarrow\pi^0\pi^0\gamma$ decay 
opens the possibility to improve our understanding of the controversial 
$I=0$, scalar resonances below and around the 1 GeV mass region. 
In our present approach we deal with these scalar resonances  by  
combining ChPT requirements with L$\sigma$M dynamics.
This treatment was first applied with some success to
$\phi\rightarrow\pi^0\eta\gamma$ \cite{Bramon:2000vu,Escribano:2000fs}
and later to $\rho,\omega\rightarrow\pi^0\pi^0\gamma$ \cite{Bramon:2001un}. 
The main feature in our approach is the use of an amplitude which agrees with ChPT
in the low part of the two-pseudoscalar invariant-mass spectrum 
but which also includes the scalar poles dominating at higher invariant-mass values. 
According to our final amplitude (\ref{AphiLsigmaM}),
the $\phi\rightarrow\pi^0\pi^0\gamma$  decay proceeds through a loop of charged kaons 
(as discussed, among others, in Refs.~\cite{Close:ay,Achasov:1987ts})
strongly coupled to the initial $\phi$ and to the final dipion system through
scalar resonance formation in the $s$-channel. 
Additional vector meson contributions, acting as a kind of background to the previous
and more interesting  scalar exchange, have been recalculated in the framework of VMD 
thus confirming older results \cite{Bramon:1992kr}.  

The measured values for the branching ratio $B(\phi\rightarrow\pi^0\pi^0\gamma)$
in Eqs.~(\ref{VEPP},\ref{KLOE}) and the shape of the $\pi\pi$ mass spectrum 
\cite{Achasov:2000ym,Akhmetshin:1999di,Aloisio:2002bt} cannot be explained without a sizable 
contribution from the $f_0$, which depends on the expression of the $f_0$ propagator.
If a simple Breit-Wigner is used, a good description of the data is achieved for
$\Gamma_{f_0}\simeq 70$ MeV and $\phi_{S}\simeq -6^\circ$.
Although these two values are only marginally consistent within our model,
the whole situation looks rather satisfactory if one takes into account the simplicity
of our approach and the well known difficulties one has to face in scalar meson physics.
If a complete one-loop expression is used for the 
$f_0$ propagator \cite{Achasov:1987ts,Escribano:2002iv},
a fully consistent description of the data is achieved with $\phi_{S}\simeq - 9^\circ$.
The range of values for $\phi_{S}$ is thus considerably restricted regardless the expression
adopted for the $f_0$ propagator. 

In turn, such small values for $|\phi_{S}|$ imply a large $\sigma\pi\pi$ coupling which,
in principle, should clearly manifest in the low part of the dipion mass spectrum.
But this is not the case in a L$\sigma $M approach like ours where the $\sigma(500)$
contribution contains a $\sigma K\bar{K}$ coupling which is proportional to 
$m^2_\sigma-m^2_K$ and thus almost vanishing. 
This depletion effect of the $\sigma(500)$ is somehow visible in the DA$\Phi$NE data 
which show very small values for $dB(\phi\rightarrow\pi^0\pi^0\gamma)/dm_{\pi\pi}$ 
in the $m_{\pi\pi}\simeq m_\sigma\simeq 500$ MeV region.
Indeed, the  chiral-loop amplitude (without scalar poles) plus the VMD background are known
to predict values \cite{Bramon:1992ki} which are two or three times larger than the data
in this range of masses.   
This feature, namely, the  prediction of small $\sigma K\bar{K}$ and large $\sigma\pi\pi$
couplings (because of their proportionality to $m^2_\sigma-m^2_K$ and to $m^2_\sigma-m^2_\pi$, 
respectively) is specific of our L$\sigma $M  approach. 
It admits a crucial and simple test.
Contrary to what happens for $m_{\pi\pi}$ around 500 MeV in
$\phi\rightarrow\pi^0\pi^0\gamma$ decays, data on $\rho\rightarrow\pi^0\pi^0\gamma$
---dominated in this case by pion rather than kaon loops--- 
should show a non-suppressed $\sigma (500)$ contribution around
$m_{\pi\pi}\simeq 500$ MeV \cite{Achasov:2002jv}. 

In summary, the recently published data on $\phi\rightarrow\pi^0\pi^0\gamma$ decays can 
be reasonably described in a rather simple context based on the complementary between 
ChPT and the L$\sigma$M.  
Future refinements of our approach and comparison with forthcoming and more accurate data
should contribute considerably to clarify  one of the most challenging aspects of present day
hadron physics, namely, the structure of the lowest lying scalar states. 
 
\section*{Acknowledgements}
Work partly supported by the EEC, TMR-CT98-0169, EURODAPHNE network.
M.~N.~was supported by CONCYTEG-Mexico under project 00-16-CONCYTEG/CONACYT-075.
J.~L.~L.~M.~acknowledges partial financial support from CONACyT and CONCyTEG.
We thank N.~N.~Achasov for useful comments.

\end{document}